\documentclass[final]{aipproc}

\layoutstyle{8x11double}

\begin{document}

\title{Charged Lepton Flavour Violation in Littlest Higgs model with
  T-parity }

\author{Naveen Gaur}{
     address={Theory Division, KEK, 1-1 Oho, Tsukuba, Ibaraki 305-0801,
     Japan.},
     email={naveen@post.kek.jp} } 


\begin{abstract}
The Little Higgs model with T-parity (LHT) belongs to the non-minimal
flavour violating model. This model has new sources of flavour and CP
violation both in quark and leptonic sectors. These new sources of
flavour violation originates by the interaction of Standard Model (SM)
fermions with heavy gauge bosons and heavy (or mirror) fermions. 
In this work we will present the impact of the new flavour structure
of T-parity models on flavour violations in leptonic sector. 
\end{abstract}

\maketitle


\section{Introduction \label{section:1}}

Little Higgs (LH) models provides a very attractive solution to the so
called {\sl little hierarchy problem} of the SM
\cite{ArkaniHamed:2001ca}. These models are 
perturbative upto the scale of $\sim 10$ TeV and have relatively smaller 
number of parameters. Unlike Supersymmetry (SUSY), in LH models the
cancellation of quadratic divergences to the Higgs mass is achieved by
the particles of the same statistics. The new particles introduced in 
these models have mass around TeV. 

In LH models, Higgs boson is kept naturally light by identifying it
with the Nambu-Goldstone Boson (NGB) of a spontaneously broken global 
symmetry and hence it remains massless at tree level. The gauge and
Yukawa interactions of NGB are introduced without generating any one
loop quadratic mass divergences in Higgs mass. This is made possible
by the mechanism called ``{\sl collective 
  breaking of symmetries}''. Two copies of the gauge group are
required to achieve the {\sl collective breaking of symmetries}. Under
{\sl collective breaking of symmetries}, the gauge and Yukawa
interactions of Higgs are introduced such that they explicitly break
the global symmetry. However the symmetry breaking is such that as
long as only one set of coupling is present, enough global symmetry is 
preserved to protect the Higgs mass. Only when both sets of
couplings are present logarithmic corrections to the Higgs mass
is generated. 

The most economical, in terms of the additional parameters and the 
particle content of the LH models, is the Littlest Higgs model
\cite{ArkaniHamed:2002qy}. In Littlest Higgs model the  
global symmetry group is SU(5) which is broken down to SO(5) at the
scale $f \sim {\cal O}(1 TeV)$. A subgroup : $[SU(2) \times U(1)]_1
\times [SU(2) \times U(1)]_2 $ of global SU(5) is gauged to provide
the gauge and Yukawa interactions in the model. In this minimal model the
additional particles which appear at TeV scale are the heavy partners
of the SM gauge bosons ($W_H, Z_H, A_H$), a heavy vector like top
quark (T) and the triplet scalar $(\Phi)$ . 

The Littlest Higgs model is very tightly constraint by the electroweak
(EW) precision tests. The reason for this is the tree level
contributions to the SM processes and triplet vev which
breaks custodial SU(2) symmetry. These constraints requires the new
physics scale to be $f \ge 3$ TeV, which re-introduces the fine tuning
and the 
{\sl little hierarchy problem}. To reconcile the LH models with EW
precision tests, Cheng \& Low \cite{Cheng:2003ju} introduced a
discrete symmetry, named {\sl T-parity} into these models. T-parity
forbids the tree level 
contributions to SM processes. The triplet vev in these model vanishes
identically and custodial symmetry is restored. Hence it is easy to
reconcile these models with the EW precision tests which brought down
the scale of these models, consistent with experimental data, to $f
\sim 500$ GeV.

\section{Lepton Flavour Violation in LHT \label{section:2}}

The detailed description of Littlest Higgs model with T-parity (LHT)
model can be found in 
\cite{Blanke:2006eb}. Here we will briefly describe the features of
LHT model required for the study of LFV. 

All the interactions in T-parity model are proposed to conserve the
discrete symmetry called T-parity. This results in the existence of a
stable T-odd particle which can be identified as a possible candidate
for dark matter.  The T-odd gauge sector of LHT consists of $W_H^\pm,
Z_H$ and a heavy photon $A_H$. Heavy photon ($A_H$) which is
electrically neutral is the lightest T-odd heavy
particle and is a suitable candidate for dark matter. 
The T-even fermionic sector of LHT consists of SM fermions and
a T-even heavy vector like top quark ($T_+$).  The T-odd fermionic
sector consists of a T-odd vector like top quark ($T_-$) and three
generation of mirror quarks and leptons, they are denoted by :
\begin{equation}
\left(\begin{array}{c}
u_H^i \\ d_H^i 
\end{array}\right), ~~
\left(\begin{array}{c}
\nu_H^i \\ \ell_H^i 
\end{array}\right) ,  \ \ \ \ {\rm with \ i = 1,2,3} 
\end{equation}
The masses of up and down type mirror fermions are equal upto the
leading order in $(v/f)$. The masses of all the new particles
introduced are of order $f$. The interaction of mirror fermions with
SM fermions and heavy gauge bosons introduces the new flavour
interactions in the 
model involving two unitary matrices in quark sector $V_{H_d}$ and
$V_{H_u}$, and two unitary matrices in leptonic sector namely
$V_{H_\nu}$ and $V_{H_\ell}$. These mirror quark and mirror lepton
matrices are related via,
$$
V^\dagger_{H_u} V_{H_d} = V_{CKM} , ~~~ V^\dagger_{H_\nu} V_{H_\ell}
= V_{PMNS} 
$$
The explicit form of these matrices are given in
\cite{Blanke:2006eb}. These new mixing matrices along with mirror
fermions are responsible for the new and rich flavour structure of LHT
model. 

The new parameters of LHT model which are relevant for the study of
LFV decays are the symmetry breaking scale ($f$), mirror lepton masses
($m_{H_i}$), mixing angles ($\theta^\ell_{ij}$) and phases
($\delta^\ell_{ij}$) of the mirror leptonic sector. These were
tabulated in \cite{Blanke:2007db} : 
$
f, \ \ m^\ell_{H_1}, \ \ m^\ell_{H_2}, \ \ m^\ell_{H_1}, \ \
\theta^\ell_{12}, \  \theta^\ell_{13}, \ \ \theta^\ell_{23}, \ \
\delta^\ell_{12}, \  \delta^\ell_{13}, \ \ \delta^\ell_{23}
$

There have been many studies of both quark and leptonic flavour sector
within the context of LH models
\cite{Buras:2006wk,Choudhury:2005jh}. Without T-parity the LH 
models are Minimal Flavour Violating (MFV) models and hence the
contribution of LH to the hadronic flavour violating processes comes
out to be small. In addition there are no new phases and hence no new
source of CP violation. In the leptonic sector of LH models due to 
triplet vev there was a possibility of writing down Lepton Number 
violating \cite{Han:2005nk,Choudhury:2005jh} interactions which then
could give rise 
to LFV. In LH model, Higgs triplet is essentially responsible for LNV
and LFV and there is no new flavour structure in the model. The
situation get changed substantially in LHT model which has additional
flavour structure. The introduction of T-parity not only makes the LH
models more consistent with EW precision tests but also give rise to
new flavour structure described by new flavour mixing matrices. This
makes the LHT model a non-MFV model which not only has a much richer
flavour structure but also has new weak phases for CP violating 
studies. Extensive studies of these in the case of quark sector has
been  done \cite{Blanke:2006eb}. In quark sector SM  
processes still play dominant role in most of the interactions
although the presence of additional weak phases can have interesting
consequences. The situation is very different in the case of leptonic 
sector. The smallness of active neutrino mass forces SM to have a
unobservably small LFV in charged lepton sector. The new flavour
structure of LHT model could provide 
much larger contribution to LFV processes which can be
observed in future experiments. The absence of QCD in leptonic sector
allows one to make very clean predictions for LFV processes. LFV
within the context of LHT model was studied in
\cite{Blanke:2007db,Choudhury:2006sq}.     

\section{Results \label{section:3}}

\begin{table}[htb]
\begin{tabular}{|c|c|c|c|}
\hline
ratio & LHT  & MSSM (dipole) & MSSM (Higgs) \\\hline 
$\frac{Br(\mu^-\to e^-e^+e^-)}{Br(\mu\to e\gamma)}$  & \hspace{.8cm}
0.4\dots2.5\hspace{.8cm}  & $\sim6\cdot10^{-3}$ &$\sim6\cdot 
10^{-3}$  \\
$\frac{Br(\tau^-\to \mu^-\mu^+\mu^-)}{Br(\tau\to \mu\gamma)}$
&0.4\dots2.3     &$\sim2\cdot10^{-3}$ & $0.06\dots0.1$ \\ 
$\frac{Br(\tau^-\to e^-\mu^+\mu^-)}{Br(\tau\to e\gamma)}$  &
0.3\dots1.6     &$\sim2\cdot10^{-3}$ & $0.02\dots0.04$ \\ 
$\frac{Br(\tau^-\to e^-e^+e^-)}{Br(\tau^-\to e^-\mu^+\mu^-)}$     &
1.3\dots1.7   &$\sim5$ & 0.3\dots0.5\\ 
$\frac{Br(\tau^-\to \mu^-\mu^+\mu^-)}{Br(\tau^-\to \mu^-e^+e^-)}$   & 
1.2\dots1.6    &$\sim0.2$ & 5\dots10 \\ \hline 
\end{tabular}
\caption{ Comparison of various ratios of branching ratios in the
  LHT model and in the MSSM using the dipole and Higgs mediated
  contributions \cite{Blanke:2007db}.}
\label{table:1}
\end{table}

LFV in LHT model was discussed for the first time in
\cite{Choudhury:2006sq} but a detailed analysis including all the LFV
processes was done in \cite{Blanke:2007db}. The estimation of 
$(g-2)_\mu$ in LHT model was done in
\cite{Blanke:2007db,Choudhury:2006sq}. It was found that muon
anomalous magnetic moment can not provide any useful constraints on
LHT parameters. 

A study of radiative LFV modes $\ell_i \to \ell_j \gamma$, where $i\ne
j$ was done in \cite{Blanke:2007db,Choudhury:2006sq}. These studies
showed that LHT can give substantial contribution to the radiative LFV
processes. In that work \cite{Choudhury:2006sq} absence of correlation
between various 
radiative decays i.e. $\mu \to e \gamma, \tau \to (\mu, e) \gamma$ was
also emphasized. This indicates that these three modes can provide
independent probes to the lepton flavour sector of the model
\cite{Choudhury:2006sq}.  
It was shown that the present limits on $\mu \to e \gamma$ could provide very
stringent constraints on LHT model, furthermore the experimental prospects
of this mode seems very promising as MEG will soon improve the 
prediction of this mode by two orders in magnitude. Although
radiative modes involving tau lepton, namely $\tau \to (\mu, e)
\gamma$, 
are not strongly correlated to $\mu \to e \gamma$, but the existing  
constraints on radiative LFV tau decays from B-factories are weak and
hence do not provide any further constraints on LHT parameter space. The
situation could change in future as SuperB factories, that are expected to
probe these decays upto the accuracy of $\sim 10^{-9}$ . 

An extensive study of the correlations of LFV processes in
LHT and its comparison with SUSY models was done in
\cite{Blanke:2007db}. A summary of their results is given in 
Table \ref{table:1}. In their work they gave the results for
LFV modes having three charged leptons in the final state {\sl i.e.}
$\mu/\tau \to \ell_i \ell_j \ell_k$ and their correlation with other
LFV modes. Their results show that the prediction of the 
ratio of the rates of $\mu^- \to e^- e^+ e^-$ to $\mu \to e \gamma$ in
LHT model can be substantially different from SUSY models. 
It is well known that in SUSY dipole operators give the
dominant contribution to these modes. 
The LFV tau decay modes, like $\tau \to \ell_i \ell_j \ell_k$ in SUSY, 
receives contributions from dipole operators and Higgs
mediated scalar operators. It is evident from the results given in
Table \ref{table:1} that in SUSY, 
Higgs mediated contributions can be dominant for modes having muons in
final state. In SUSY the modes dominated by dipole 
contributions show very strong  correlation between $\tau \to \ell_i
\ell_j \ell_k$ and $\tau \to \ell_i \gamma$ (with i = e,$\mu$) which
tends to get relaxed for Higgs mediated contributions. But the
predictions of these ratios in LHT are strikingly different from SUSY
models even if we include Higgs contributions.   

Another notable difference in SUSY and LHT model predictions of LFV
processes comes while correlating processes $\tau \to \ell_i \ell_j
\ell_k$ {\sl i.e.} tau decays having three leptons in final state. 
For this purpose following ratios were constructed in 
\cite{Blanke:2007db} : 
$$
R_1 = \frac{Br(\mu^- \to e^- e^+ e^-)}{Br(\mu \to e \gamma)} ,
$$
$$
R_2 = \frac{Br(\tau^- \to e^- e^+ e^-)}{Br(\tau^- \to e^- \mu^+ \mu^-)} , ~~
R_3 = \frac{Br(\tau^- \to \mu^- \mu^+ \mu^-)}{Br(\tau^- \to \mu^- e^+
  e^-)}  
$$

As can be seen from table \ref{table:1} these ratios can be substantially
different in SUSY and LHT. 
The reason for these differences in LHT and SUSY lies in the mechanism
responsible for LFV in these models. Whereas dipole operators are
responsible for radiative modes ($\ell_i \to \ell_j \gamma$)  in both
these models, in SUSY, LFV in $\mu/\tau \to
\ell_i \ell_j \ell_k$ arises due to photon mediated dipole operators and
Higgs mediated scalar operators. The Higgs mediated scalar operators
can be dominant in decays involving $\tau$ lepton. In SUSY 
both of these contributions come from penguin or self energy
diagrams. On the other hand in LHT the
dipole contributions can be almost completely neglected in comparison
with the Z-penguin and box diagrams. 
The dipole dominance gives a relatively stable prediction for $R_1$ in
SUSY whereas this ratio can change a bit in LHT where $\mu^- \to
e^- e^+ e^-$ is not dominated by dipole contributions. 
$R_2$ and $R_3$ for LHT model are of order one and do not change much
on changing the model parameters. This is because in LHT the Z-penguin
and box diagram contributions are nearly equal for the decays of type, 
$\tau \to \ell_i \ell_j \ell_k$ and hence the ratios of these processes
is stable. In the case of SUSY, Higgs mediated diagrams give rise to
scalar operators that can alter the predictions of the mode having
muons in final state ($\tau^- \to \mu^- \mu^+ \mu^-$) as opposed to
LHT.   

In summary, LHT model has the structure to provide LFV which can be
observed in future experiments. LFV processes, if observed, can also
be used to distinguish the models responsible for these processes.  


\begin{theacknowledgments}
I thank the other authors of \cite{Choudhury:2006sq} : S. Rai Choudhury,
A. S. Cornell, A. Deandrea and Ashok Goyal. 
I would also like to thanks Andrzej J. Buras for discussion in regard
to LFV in LHT models. 
This work is supported by JSPS grant no. P-06043. 
\end{theacknowledgments}



\begin{thebibliography}{99}

\bibitem{ArkaniHamed:2001ca}
  N.~Arkani-Hamed, A.~G.~Cohen and H.~Georgi,
  Phys.\ Rev.\ Lett.\  {\bf 86}, 4757 (2001)
, 
%
  N.~Arkani-Hamed, A.~G.~Cohen and H.~Georgi,
  Phys.\ Lett.\  B {\bf 513}, 232 (2001)
.


\bibitem{ArkaniHamed:2002qy}
  N.~Arkani-Hamed, A.~G.~Cohen, E.~Katz and A.~E.~Nelson,
  JHEP {\bf 0207}, 034 (2002)
,
%
  T.~Han, H.~E.~Logan, B.~McElrath and L.~T.~Wang,
  Phys.\ Rev.\  D {\bf 67}, 095004 (2003)
.


\bibitem{Cheng:2003ju}
  H.~C.~Cheng and I.~Low,
  JHEP {\bf 0309}, 051 (2003)
,
%
  H.~C.~Cheng and I.~Low,
  JHEP {\bf 0408}, 061 (2004)
.


\bibitem{Blanke:2006eb}
  M.~Blanke, A.~J.~Buras, A.~Poschenrieder, S.~Recksiegel, C.~Tarantino, S.~Uhlig and A.~Weiler,
  JHEP {\bf 0701}, 066 (2007)
,
%
  J.~Hubisz, S.~J.~Lee and G.~Paz,
  JHEP {\bf 0606}, 041 (2006)
,
%
  M.~Blanke, A.~J.~Buras, S.~Recksiegel, C.~Tarantino and S.~Uhlig,
  JHEP {\bf 0706}, 082 (2007)
,
%
  M.~Blanke, A.~J.~Buras, S.~Recksiegel, C.~Tarantino and S.~Uhlig,
  arXiv:hep-ph/0703254 ,
%
  M.~Blanke, A.~J.~Buras, A.~Poschenrieder, S.~Recksiegel, C.~Tarantino, S.~Uhlig and A.~Weiler,
  Phys.\ Lett.\  B {\bf 646}, 253 (2007), 
%
  M.~Blanke, A.~J.~Buras, A.~Poschenrieder, C.~Tarantino, S.~Uhlig and A.~Weiler,
  JHEP {\bf 0612}, 003 (2006).


\bibitem{Buras:2006wk}
  A.~J.~Buras, A.~Poschenrieder, S.~Uhlig and W.~A.~Bardeen,
  JHEP {\bf 0611}, 062 (2006)
,
%
  A.~J.~Buras, A.~Poschenrieder and S.~Uhlig,
  Nucl.\ Phys.\  B {\bf 716}, 173 (2005), 
%
  S.~R.~Choudhury, N.~Gaur, A.~Goyal and N.~Mahajan,
  Phys.\ Lett.\  B {\bf 601}, 164 (2004)
,
%
  S.~R.~Choudhury, N.~Gaur, G.~C.~Joshi and B.~H.~J.~McKellar,
  arXiv:hep-ph/0408125.


\bibitem{Han:2005nk}
  T.~Han, H.~E.~Logan, B.~Mukhopadhyaya and R.~Srikanth,
  Phys.\ Rev.\  D {\bf 72}, 053007 (2005)
,

\bibitem{Choudhury:2005jh}
  S.~R.~Choudhury, N.~Gaur and A.~Goyal,
  Phys.\ Rev.\  D {\bf 72}, 097702 (2005)
,
%
  A.~Goyal,
  Mod.\ Phys.\ Lett.\  A {\bf 21} (2006) 1931 , 
%
  C.~X.~Yue and S.~Zhao,
  Eur.\ Phys.\ J.\  C {\bf 50}, 897 (2007),


\bibitem{Blanke:2007db}
  M.~Blanke, A.~J.~Buras, B.~Duling, A.~Poschenrieder and C.~Tarantino,
  JHEP {\bf 0705}, 013 (2007).


\bibitem{Choudhury:2006sq}
  S.~R.~Choudhury, A.~S.~Cornell, A.~Deandrea, N.~Gaur and A.~Goyal,
  Phys.\ Rev.\  D {\bf 75}, 055011 (2007),
%
  A.~Goyal,
  arXiv:hep-ph/0609095.


\end{thebibliography}
\end{document}